
\input phyzzx
\nopagenumbers

\Pubnum={IASSNS-HEP-92/46\cr CfPA-TH-92/17\cr NSF-ITP-92-120\cr UMAC92-1}
\date={June 1992}
\titlepage
\title{Instability and Subsequent Evolution of Electroweak Bubbles}
\author{Marc Kamionkowski\foot{e-mail: kamion@guinness.ias.edu.}\foot{SSC
Fellow}}
\address{School of Natural Sciences, Institute for Advanced
Study, Princeton, NJ 08540}
\andaddress{Center for Particle Astrophysics, University of California,
Berkeley, CA 94720}
\andauthor{Katherine Freese\foot{e-mail:
freese@umiphys.bitnet}\foot{Presidential Young Investigator and Sloan
Foundation Fellow}}
\address{Department of Physics, University of Michigan,
Ann Arbor, MI 48109}
\andaddress{Institute for Theoretical Physics,
University of California, Santa Barbara, CA 93109}
\abstract
\singlespace

Bubbles in a first-order electroweak phase transition are nucleated with
radii $R_0$ and expand with velocity $v$. If $v$ is subsonic, a bubble
becomes unstable to  non-spherical perturbations when its radius is
roughly $10^4\, R_0$.  These perturbations accelerate the transition, and
the effective velocity of bubble growth rapidly becomes supersonic.  The
transition should  subsequently proceed spherically via detonation.  If
for some reason the onset of detonation is postponed, the surface area of
the bubbles may be enhanced by $10^5$.  We discuss consequences for
electroweak baryogenesis.

\endpage
\baselineskip=18pt
\overfullrule=0pt
\pagenumber=2
\pagenumbers

\def\fun#1#2{\lower3.6pt\vbox{\baselineskip0pt\lineskip.9pt
  \ialign{$\mathsurround=0pt#1\hfil##\hfil$\crcr#2\crcr\sim\crcr}}}
\def\lap{\mathrel{\mathpalette\fun <}}
\def\gap{\mathrel{\mathpalette\fun >}}

\def\etal{{\it et al.}}

\def\Mp{M_{Pl}}

\REF\cronin{J. H. Christenson \etal, \sl Phys. Rev. Lett. \bf
13\rm, 138 (1964).}
\REF\thooft{G. 't Hooft, \sl Phys. Rev. Lett. \bf 37\rm,1154 (1976).}
\REF\sphaleron{S. Dimopoulos and L. Susskind, \sl Phys. Rev. D \bf 18\rm,
4500 (1978); V.~A.~Kuzmin, V.~A.~Rubakov, M.~E.~Shaposhnikov, \sl Phys.
Lett. \bf 155B\rm, 36 (1985); N.~Manton, \sl Phys. Rev. D \bf 18\rm, 4500
(1978); F.~Klinkhammer and N.~S.~Manton, \sl Phys. Rev. D \bf 30\rm, 2212
(1984).}
\REF\sakharov{A. D. Sakharov, \sl JETP Letters, \bf 5\rm, 24 (1967).}
\REF\dine{M. Dine \etal, SLAC-PUB-5741 (1992); B.~H.~Liu,
L.~Mclerran, and N.~Turok,
TPI-MINN-92/18-T (1992); S. Y. Khlebnikov, UCLA/92/TEP/20 (1992).}
\REF\landau{L. Landau and D. Lifshitz, \it Fluid Mechanics \rm (Pergamon,
New York, 1959).}
\REF\zeldovich{Ya. B. Zeldovich \etal, \it The Mathematical
Theory of Combustion and Explosions \rm (Plenum, New York, 1985).}
\REF\link{B. Link, \sl Phys. Rev. Lett. \bf 68\rm, 2425 (1992);
K.~Kajantie, NSF-ITP-92-62 (1992).}
\REF\enqvist{K. Enqvist \etal, \sl Phys. Rev. D \bf 45\rm, 3415 (1992).}
\REF\gyulassy{M. Gyulassy \etal,
\sl Nucl. Phys. \bf B237\rm, 477 (1984); L.~van~Hove,
\sl Z.~Phys. C, \bf 21\rm, 93 (1983).}
\REF\turok{N. Turok, \sl Phys. Rev. Lett. \bf 68\rm, 1803 (1992).}
\REF\woosley{S. E. Woosley in \it Particle Astrophysics: Forefront
Experimental Issues, \rm Proceedings of the Workshop, Dec. 8-10, 1988,
Berkeley, CA, edited by E.~B.~Norman (World Scientific,
Singapore, 1989).}
\REF\mandelbrot{B. B. Mandelbrot, \it The Fractal Geometry of
Nature, \rm (W.~H.~Freeman and Co., New York, 1983).}
\REF\steinhardt{P. J. Steinhardt, \sl Phys. Rev. D \bf 25\rm, 2082
(1982).}
\REF\khokhlov{A. M. Khokhlov, \sl Astron. Astrophys. \bf 246\rm,
383 (1991).}
\REF\texture{N. Turok and J. Zadrozny, \sl Phys. Rev. Lett. \bf 65\rm,
2331 (1990); \sl Nucl. Phys. \bf B358\rm, 471 (1991);
L.~McLerran \etal, \sl Phys. Lett. \bf 256B\rm,
451 (1991).}
\REF\cohen{A. E. Nelson, D. B. Kaplan, and A. G. Cohen, UCSD/PTH 91-20
(1991); \sl Phys. Lett. \bf 263B\rm, 86 (1991); \sl Nucl. Phys. \bf
B349\rm, 727 (1991).}
\REF\freese{K. Freese and F. C. Adams, \sl Phys. Rev. D \bf 41\rm, 2449
(1990).}

\unnumberedchapters

There has been much interest in the dynamics of a possible first-order
electroweak phase transition (EWPT) recently.  The motivation is clear:
one of the fundamental problems in particle physics and cosmology is the
origin of the baryon asymmetry of the Universe (BAU).  While
$C$ and $CP$ violation [\cronin] as well as baryon-number violating
instanton effects [\thooft] have been known to exist in the
Standard Model for quite some time, only recently has it been suggested
that the rate for baryon-number violating interactions may become
appreciable at high temperatures [\sphaleron].  If the
EWPT is first order, the third of
Sakharov's [\sakharov] criteria for baryogenesis,
out-of-equilibrium processes, may also be found in electroweak (EW)
physics.  Thus, the BAU may be explained in the SSC era.

In the standard picture of a first order EWPT,
spherical bubbles are nucleated with
microphysical radii $R_0 \sim 10^{-17}$ cm
and then expand spherically with velocity $v$
to macroscopic radii $R_{perc}\sim 10^{14}\,v R_0$ before they collide.
The bubble-wall velocity $v$ is still uncertain, but
recent estimates suggest that a wall
may propagate subsonically (\ie, as a deflagration front) [\dine].
In this Letter we show that shape instability of the bubble wall
rapidly causes the propagation to become turbulent
and proceed more quickly, and probably instigates the
onset of detonation; then the bubbles
expand spherically to fill space at supersonic velocities.

A deflagration front is unstable to perturbations
with wavelengths in the range $\lambda_c\lap\lambda\lap\lambda_{max}$
where $\lambda_c\sim R_0 v^{-2}$ is set by the
surface tension in the wall [\landau,\zeldovich,\link], and
$\lambda_{max}$, which is proportional to the
radius $R$ of the bubble, is set by the underlying
expansion of the bubble [\zeldovich].  After
the bubbles are nucleated, they expand spherically until $\lambda_{max}$
reaches $\lambda_c$; then $R_{inst}\sim100\,
v^{-2} R_0$ and hydrodynamic instabilities set in.  The
subsequent bubble shape will be roughly spherical;
however, instead of a smooth surface, the wall
is highly wrinkled
with distortions that enhance
the  surface area of the wall and thereby accelerate
the transition.  Although the details are far from understood
(in any fluid dynamic system),
the onset of
turbulence and corresponding acceleration of the transition should result
in a detonation front shortly after the bubbles are nucleated, when
roughly a fraction $10^{-21}$ of the Universe has been converted
to the low-temperature phase.  On the
other hand, if for some reason the transition continues  as a deflagration,
then the surface area of the walls is enhanced by five orders of magnitude
by the time the bubbles percolate.  In either case, the dynamics of the
transition assumed in models where baryon number is produced in a
first-order EWPT could be significantly altered.

We limit ourselves to the case where the
transition occurs at
temperatures near the critical temperature $T_c$,
the latent heat is small compared to the thermal energy density, and
$v\ll1$; under
these assumptions the calculations simplify considerably.
Such conditions are possible in the Standard Model [\dine].  We feel that
a more general analysis should result in similar conclusions.
We neglect the expansion of the
Universe, since the timescale for the EWPT is much
smaller than the expansion timescale [\dine,\enqvist].

First we review some results from the theory of
combustion of relativistic fluids [\gyulassy] in the case where fluid
velocities are nonrelativistic.
Consider a planar interface in the $y$-$z$ plane that
propagates in the $-x$ direction.  Then in the rest frame
of the wall, matter in the
symmetric phase
enters  the interface with a velocity $v_1$, and matter in the
broken-symmetry phase
leaves the interface with velocity $v_2$.
Conservation of energy and momentum across the
interface leads to the conditions
$$
w_1 v_1 = w_2 v_2, \quad {\rm and} \quad p_1 = p_2,
\eqn\firstbc
$$
where $w=e+p$ is the enthalpy density, $e$ is the energy
density, and $p$ is the pressure.  Throughout, the subscript
``1'' refers to the symmetric phase and the subscript ``2'' refers to the
broken-symmetry phase.

The $e_i$ and $p_i$ may be
obtained from finite-temperature field theory. Near
$T_c$ the effective potential may
be written [\dine,\enqvist,\turok],
$$
V(\phi,T)=(1/2)\gamma(T^2-T_0^2)\phi^2 - (1/3) \alpha T
\phi^3 + (1/4)\lambda \phi^4,
\eqn\effectivepot
$$
where $\phi$ is the Higgs field, $T_0^2=(\gamma/\lambda)\phi_0$,
$\phi_0=250$ GeV is the Higgs vacuum expectation value, and
$\alpha$, $\lambda$, and $\gamma$ are parameters that depend on
the $W$, $Z$, and top-quark masses, and on the Higgs structure
of the theory.  The critical temperature $T_c$ is defined as the
temperature at which there exists a second minimum of $V$
degenerate with the minimum at the origin; for the
effective potential above, $ T_c^2 = T_0^2/ (
1-{2\over9}{\alpha^2\over \gamma\lambda})$ [\enqvist].

Although the difference in free energies $B(T)$ between the
two phases is in general a complicated function
of $T$, if the transition occurs near $T_c$
then
$ B(T) \simeq(L/4)[1-(T^4 / T_c^4)]$, where $ L= -T_c (\partial B/
\partial T)|_{T_c} = (4\alpha^2\gamma/ 9\lambda^2) T_0^2 T_c^2$ is the
latent heat of the
transition [\enqvist].  This
leads to the rather simple equation of state (which mimics the
QCD bag model),
$$
p_1(T)=[w_1(T)-L]/4, \quad {\rm and} \quad
e_1(T)= [3 w_1(T) + L]/4,
\eqn\higheos
$$
$$
p_2(T) = w_2(T)/4, \quad {\rm and} \quad
e_2(T) =  3 w_2(T)/4,
\eqn\loweos
$$
where $w_1(T)=a_1 T^4$, and $w_2(T)=a_2 T^4$, $a_1 \simeq a_2 \sim 100$
and $a_1 -a_2=L(4T_c^4)^{-1}$ (see Ref.~\enqvist).

We now study the hydrodynamic stability of a
spherical bubble to small non-spherical perturbations.  For
distortions with $\lambda\ll R$, it is valid to treat the wall as a
planar interface.  The stability of a planar front for combustion of a
relativistic gas was recently discussed by Link [\link].
Consider a small perturbation to the planar discontinuity
of the form $x_f=d\exp(iky+\omega t)$.  This disortion in
the wall surface will be accompanied by perturbations to the velocity and
pressure.  If $\vec v$ and $p$ are unperturbed quantities,
then the perturbations $p'$ and ${\vec v}'$ must satisfy [\link]
$$
\left[ {\partial\over\partial t} +\vec v \cdot
\vec \nabla \right] p' +w c_s^2
\vec \nabla \cdot {\vec v}' =0,\quad {\rm and} \quad \left[
{\partial\over\partial t} +\vec v \cdot \vec \nabla \right]
{\vec v}' + {1\over w}\vec \nabla p'=0,
\eqn\hydroone
$$
where $c_s=1/\sqrt{3}$ is the speed of sound.
There are four boundary conditions on the discontinuity that
the perturbed quantities must satisfy.  The first,
$$
p_1'=p_2'-\sigma\left({\partial^2\over\partial y^2} -
{\partial^2 \over \partial t^2}\right)x_f,
\eqn\bcone
$$
follows from Eq.~\firstbc\ and includes the effects of
surface tension and finite mass density of the wall; these two effects
favor a flat surface.
The boundary condition on $(w v_x)'$ (to
lowest order in $v_1$ and $v_2$,), from Eq.~\firstbc, is
$$
w_1\left(v_{x1}'-\partial x_f/\partial t\right)
= w_2\left(v_{x2}'-\partial x_f/\partial t \right).
\eqn\bctwo
$$
Requiring that the tangential velocities on both sides be
equal leads to
$$
v_{y1}'+v_1 (\partial x_f / \partial y) = v_{y2}' + v_2
(\partial x_f / \partial y).
\eqn\bcthree
$$
We make the {\it ansatz} that the enthalpy flux across
the interface is proportional to the net blackbody energy flux
across the interface.  Then the perturbed enthalpy
flux is [\link]
$$
(wv)'= (3/4)\alpha(p_1'-p_2'),
\eqn\finalbc
$$
where $\alpha$ is a fudge factor, $0\leq\alpha\leq1$, and
$\alpha=1$ in case the enthalpy flux {\it is} the blackbody
energy flux.  (Our final results will not depend
on $\alpha$.)  Equating Eqs.~\bctwo\ and \finalbc\ gives us our
fourth boundary condition.

If nontrivial solutions that satisfy Eq.~\hydroone\ and the boundary
conditions can be found for some $\omega>0$, then there are growing
modes, and the wall is unstable to small perturbations.
To satisfy the equations of motion and the boundary conditions,
$\omega$ must satisfy [\link]
$$
\omega^2(v_1+v_2)+2\omega v_1 v_2 + \left[ k^2(v_1 -v_2) +
{\sigma k^3 \over w_1 v_1} \right] v_1 v_2 =0.
\eqn\dispersion
$$
If $v_2>v_1$ (\ie, if the phase transition proceeds via
deflagration), then Eq.~\dispersion\ has a positive root
for wave numbers $k<k_c=(v_2-v_1)w_1 v_1/\sigma$, and there are growing
modes.  For larger wave numbers the system is stabilized by surface
tension.

{}From Eqs.~\firstbc, \higheos, and \loweos, we find $v_2-v_1 \simeq
(L/w_1) v_1$; thus
the wall is unstable to small perturbations with
$$
\lambda\gap\lambda_c\equiv k_c^{-1}\simeq \sigma / (L v_1^2).
\eqn\lambdasmall
$$
In the limit of small supercooling
the surface tension is given by $ \sigma=(2^{3/2}\alpha^3 /
3^4 \lambda^{5/2}) T_c^3$ [\enqvist].  Determining $v_1$ is much
more difficult and requires an investigation of the microscopic
interactions of the particles in the thermal bath with the
advancing wall.  Recent estimates suggest that for the minimal
standard model, the wall velocity may
be in the range $0.01\lap v_1\lap 0.3$ [\dine].
Note that $\lambda_c\sim R_0 v_1^{-2} \sim 10^{-15}$ cm (for $v_1\sim
0.1$).

As long as $\lambda\ll R$, the analysis assuming
a planar interface should be valid; however, for $\lambda \sim R$,
one should take the expansion of the bubble into account.  As the
bubble expands, the wavelength of the perturbation increases.
If the amplitude of a perturbation grows more slowly than the
wavelength, then the distortion
is smoothed out in time.  For our case of a weak
transition, $\delta\equiv(v_1-v_2)/v_1 \simeq L/w \sim 0.01\ll 1$, the
growth rate for an instability with a large $\lambda$
is found from Eq.~\dispersion\ to be
$\omega\simeq \delta v_1 k/2$, while the growth rate for the
bubble is roughly $v_1/R$. For a perturbation to be
to be unstable it must have $k\gap2/(R\delta)$, while
perturbations with $\lambda\gap\lambda_{max}\simeq
R\delta/2$ will be stabilized.  Although this derivation is
heuristic, for $\delta\ll 1$ it reproduces the results
of the exact analysis for the case of spherical combustion of a
nonrelativistic gas, and it should be a good
approximation in the case of a slowly-moving relativistic gas
as considered here.  So, perturbations with
$\lambda\lap\lambda_c$ will be stabilized by surface tension, and
those with $\lambda\gap\lambda_{max}$ will be
stabilized by the growth of the bubble.

In the standard picture of a first-order EWPT, bubbles
are nucleated with radii $R_0\sim\sigma/L$ and then grow spherically with
velocities $v$ until the bubbles percolate.  The radius of the bubble at
this time is (again, assuming typical parameters [\enqvist]) $R_{perc}
\sim 10^{-4}\,v (\Mp/T_c) T_c^{-1}\sim 10^{14}\,v R_0$.  On the other
hand, when $\lambda_{max}$ reaches $\lambda_c$,
$R_{inst} \simeq \sigma w/(L^2 v^2) \simeq 100 v^{-2} R_0$; at this
point the bubble becomes unstable to non-spherical perturbations.  Since
$R_{perc}$ is many orders of magnitude larger than $R_{inst}$ the
perturbations have plenty of time to mature and the standard picture of
bubble evolution may be drastically altered.

If the bubble volume is $V$, the bubble will look somewhat spherical with
a nominal radius $R$ given by $V=(4/3)\pi R^3$; however,
instead of a smooth surface, the
wall is highly wrinkled on scales $\lambda_c\lap\lambda\lap
\lambda_{max}$, and the surface area of such a bubble
is actually much larger than $4\pi R^2$.
Perturbations to the fluid flow accompany those in the bubble
surface, so that the normal flow velocity of fluid across the interface
is $v_1$ at every point on the surface [\cf\ Eqs.~\bctwo\
and \finalbc]; therefore, the rate of the transition is enhanced.
A similar situation
arises in supernova theory, where the burning of a carbon-oxygen
white dwarf proceeds via deflagration and the rate at which
burning occurs is proportional to the surface area of the
wrinkled flame [\woosley].
The surface area is enhanced roughly by a factor
$$
{{\rm surface\, area}\over 4 \pi R^2} \simeq \left( { \lambda_{max} \over
\lambda_c } \right)^{D-2},
\eqn\enhancement
$$
where the fractal
dimension $D$ [\mandelbrot] is some number between 2 and 3
but most likely near 2.6 [\woosley]. The effective
velocity
$v_{eff}\equiv (dR/dt)$ at which
a sphere of comparable volume as the bubble would expand becomes
$$
v_{eff}\simeq\left(\lambda_{max} / \lambda_c\right)^{D-2} v_2.
\eqn\veff
$$
Although the exact fractal dimension is uncertain, the
qualitative form of Eq.~\veff\ is correct; $v_{eff}$
might differ from our estimate by an order of magnitude or
so, but this has little effect on our conclusions.
There is a possibility that once the
perturbation goes nonlinear (\ie, its amplitude becomes
comparable to its wavelength) that it becomes stabilized and the
resulting flow is not turbulent [\zeldovich].  However, in
this case, the wall would still be wrinkly on
length scales from $\lambda_c$ to $\lambda_{max}$.  The resulting
surface-area enhancement and $v_{eff}$ would still be
comparable to
those given in Eqs.~\enhancement\ and \veff.  Shortly
after instabilities set in, the transition accelerates and when
$R\sim\lambda_c \delta^{-1}
v_1^{1/(2-D)} \sim R_0 v^{-4} \delta^{-1} \ll R_{perc}$,
$v_{eff}$ becomes supersonic.
At this point only a fraction $(R/R_{perc})^3\sim10^{-21}$ (for
$v_1\sim0.1$) of the Universe has been converted to the new
phase.  Therefore, if baryogenesis occurs at the EWPT, it
takes place after $v_{eff}$ becomes supersonic.

The most likely scenario is that when $v_{eff}$ increases past
$c_s$, a detonation wave sets in.  Simply stated,
the reason is that the deflagration front is preceded by a fluid
flow, and it is hard to see how the appropriate fluid flow can
be maintained in front of a deflagration wave itself
moving supersonically.  In the frame of the deflagration front,
the flow velocity of fluid
into the interface is smaller than the flow velocity of fluid out of the
interface (\ie, $v_2>v_1$).
In the ``rest" frame of the Universe, the fluid
is at rest far away from the transition;
furthermore, by symmetry arguments, the fluid
inside the bubble must be
at rest.  There is a piston effect
as the wall pushes the fluid outside the bubble with a speed
$v_2 -v_1 = \delta v_1$.  A precompression shock precedes the
deflagration front and accelerates the fluid, which is
initially at rest, radially outward to a velocity
$\delta v_1$ [\gyulassy].  If the wall is distorted and the
transition is accelerated, the wall pushes the fluid outside the
bubble with a velocity near $\delta v_{eff}$.  Since $\delta
v_{eff}\ll1$, only a weak shock is needed.  Weak shocks travel at
velocities only slightly larger than $c_s$ [\steinhardt];
once $v_{eff}\gap c_s$, the deflagration front
will merge with the shock to form a detonation
wave.  Although the exact mechanism for onset of detonation from
deflagration is still under investigation in fluid
systems and is not entirely understood even in the nonrelativistic
case [\zeldovich,\khokhlov], the onset of
detonation from a shock preceding an accelerating turbulent deflagration
front is observed in laboratory experiments [\zeldovich], and
appears in the theory of Type Ia
supernovae [\khokhlov].  In order to satisfy the hydrodynamical
equations of motion with the boundary conditions that the fluid far from
the bubble as well as at the center of the bubble be at rest,
the Chapman-Jouget condition must be satisfied, and the bubble expands at
a velocity $v\simeq \sqrt{1/3} + \sqrt{2\delta/9}$ (for $\delta \ll
1$) slightly larger than $c_s$ [\steinhardt].
Since perturbations cannot propagate faster than $c_s$,
perturbations should be smoothed out and the bubble
expands spherically to fill all space.

A detonation wave heats the gas as it passes, so one might worry that if
the gas is heated to a temperature above $T_c$ that the phase transition
cannot continue.
A detonation will certainly propagate if the supercooling of the
Universe is greater than the subsequent heating, as is found in some
models (though not all).
Further work should investigate the details of the phase
transition at a detonation front in the case that the gas is heated
above $T_c$.

On the other hand, if for some unforeseen reason,
a terminal $v_{eff}$ smaller than $c_s$ is
reached, then the subsequent evolution could continue
as a deflagration with a distorted surface.  This distortion can be quite
dramatic:  By the time the bubbles percolate, the surface
area of the bubbles is enhanced by roughly $(R_{perc}\delta/\lambda_c
)^{D-2} \sim10^5$ [\cf, Eq.~\enhancement].  If this is
the case it might play a role in EW baryogenesis.  However, the
baryon number in recently proposed existing models
where baryogenesis occurs at the phase boundary [\texture,\cohen]
should be unaltered.  Although the transition in this case would be
accelerated, the resultant baryon number is generally proportional to the
amount of fluid that passes through
the wall and this remains unaltered by
turbulence.  In some models such as that in Ref.~\cohen, the rate of
baryogenesis depends on transport of particles
near the wall and the
final baryon number depends on the wall velocity; in such models,
the resulting baryon asymmetry depends not on $v_{eff}$ but
only on the {\it local} velocity of fluid through the
wall, which remains unchanged (in the nonrelativistic limit) by
turbulence.
We can only speculate that relativistic corrections could actually alter
the flow velocity across the wall.  Another possible effect of the wall
convolution is that in models where transport near the wall is crucial,
particles could multiple
scatter off one wall into another wall; however, this would
require that $\lambda_c$ be smaller
than the particle mean-free path.

Throughout we assumed that latent
heat is transported from the surface hydrodynamically.  If, on
the other hand, radiative transport is important and bubble
growth is limited by diffusion of latent heat from the
wall, then the wall may become unstable on length scales larger than
mean-free path of radiation as shown by Freese and
Adams [\freese] for the case of a first-order QCD phase
transition.  If so, instabilities may set in even
earlier than we found (as soon as the bubbles nucleates), and the
surface-area enhancement could possibly be even larger than we
estimated.  The resulting bubble shape in this case may deviate
drastically from spherical; the bubble looks like seaweed.  Multiple
scatter would also become more important in this case.

To summarize, the propagation of
a deflagration front in a weakly first order EWPT
becomes turbulent, the
transition is accelerated, and the effective propagation velocity
of the walls rapidly becomes supersonic.  Under these
conditions the deflagration front could turn into a detonation
shortly after the bubble is nucleated, and the macroscopic
growth of the bubbles should occur via a detonation wave
traveling near $c_s$.  Our results suggest that,
due to hydrodynamic effects, macroscopic bubble propagation may be
significantly different from what one would expect from detailed
studies of the microscopic
kinetics [\dine].  This should
come as no surprise; it has long been known that the propagation
velocity of a spherical detonation wave is determined by
hydrodynamics (the Chapman-Jouget
condition [\landau,\gyulassy,\steinhardt]) and not by the
microscopic kinetics of the reaction.

Strictly speaking, our analysis is
valid only for nonrelativistic propagation velocities and for transitions
with small latent heat, but a more general analysis
under less restrictive assumptions should result in similar conclusions.
For example, as long as the
deflagration velocity $v$ is subsonic, small perturbations could
propagate ahead of the detonation front, and the hydrodynamic
instability should exist; in addition, as $v$ is
increased (while still subsonic),
$\lambda_c$ is decreased [\cf, Eq.~\lambdasmall] so the instability
should set in sooner.  If the latent heat is increased,
$\lambda_c$ decreases and $\lambda_{max}$ becomes larger,
so turbulence should set in sooner.   Also, for larger latent heats,
the detonation front propagates at a larger
velocity [\steinhardt].  The transition from deflagration to
detonation may also be important for the dynamics of the QCD phase
transition if it is first order.

We thank A. Burrows, G. Fuller, P. Pinto, and K.
Rajagopal for useful discussions.
MK was supported in part by the Texas National Research
Laboratory Commission. KF was supported in part by a
Presidential Young Investigator award, a Sloan Foundation
Fellowship, and by NSF grant
NSF-PHY-92-96020.  MK acknowledges the hospitality
of the Center for Particle
Astrophysics, and Lawrence Berkeley Laboratory, and  KF
the hospitality of the ITP at Santa Barbara and the Aspen
Center for Physics.

\refout
\endpage
\end